\documentclass[aos]{imsart}

\RequirePackage{amsthm,amsmath,amsfonts,amssymb,ulem}
\RequirePackage[numbers]{natbib}
\RequirePackage{graphicx}

\graphicspath{{figures/}}

\def\Xb{{\boldsymbol X}}

\def\Zb{{\boldsymbol Z}}

\def\zb{{\boldsymbol z}}

\usepackage{color}

\renewcommand{\bar}{\overline}

\def\vec{{\rm vec}}

\def\1{{\bm 1}}
\def\0{{\bm 0}}
\def\th{{\rm th}}

\def\rmspan{{\rm span}}

\def\IF{{\rm IF}^{(1)}}

\def\Sr{\mathcal{S}_r}
\newcommand{\real}{\mathbb{R}}
\renewcommand{\tilde}{\widetilde}
\renewcommand{\hat}{\widehat}

\newtheorem{thm}{Theorem}

\newtheorem{lem}{Lemma}

\newtheorem{rmk}{Remark}
\newtheorem{cor}[thm]{Corollary}
\newtheorem{ex}{Example}[section]

\startlocaldefs

\endlocaldefs

\begin{document}

\begin{frontmatter}
\title{On the efficiency-loss free ordering-robustness of product-PCA}
\runtitle{Ordering-robustness of product-PCA}

\begin{aug}
\author[A]{\fnms{Hung}~\snm{Hung}\ead[label=e1]{hhung@ntu.edu.tw}}
\and
\author[B]{\fnms{Su-Yun}~\snm{Huang}\ead[label=e2]{syhuang@stat.sinica.edu.tw }}
\address[A]{Institute of Health Data Analytics and Statistics,
National Taiwan University, Taiwan\printead[presep={,\ }]{e1}}

\address[B]{Institute of Statistical Science, Academia Sinica, Taiwan\printead[presep={,\ }]{e2}}
\end{aug}

\begin{abstract}
This article studies the robustness of the eigenvalue ordering, an important issue when estimating
the leading eigen-subspace by principal component analysis (PCA). In Yata and Aoshima (2010), cross-data-matrix PCA (CDM-PCA) was proposed and shown to have smaller bias than PCA in estimating eigenvalues. While CDM-PCA has the potential to achieve better estimation of the leading eigen-subspace than the usual PCA, its robustness is not well recognized. In this article, we first develop a more stable variant of CDM-PCA, which we call product-PCA (PPCA), that provides a more convenient formulation for theoretical investigation. Secondly, we prove that, in the presence of outliers, PPCA is more robust than PCA in maintaining the correct ordering of leading eigenvalues. The robustness gain in PPCA comes from the random data partition, and it does not rely on a data down-weighting scheme as most robust statistical methods do. This enables us to establish the surprising finding that, when there are no outliers, PPCA and PCA share the same asymptotic distribution. That is, the robustness gain of PPCA in estimating the leading eigen-subspace has no efficiency loss in comparison with PCA. Simulation studies and a face data example are  presented to show the merits of PPCA. In conclusion, PPCA has the potential to supplement the role of the usual PCA in real applications whether outliers are present or not.
\end{abstract}



\end{frontmatter}


\section{Introduction}\label{sec.introduction}

Principal component analysis (PCA) is the most widely used linear dimension reduction method (Jolliffe and Cadima, 2016). It aims to search lower dimensional combinations of covariates while preserving the variation of the data as much as possible.
Let $X\in \mathbb{R}^p$ be a random vector generated from the cdf $F$ with mean $\mu$ and covariance $\Sigma$. For simplicity and without loss of generality, assume $\mu=0$. Let the eigenvalue decomposition (EVD) of $\Sigma$ be
\begin{eqnarray}
\Sigma&=&\Gamma\Lambda\Gamma^\top,
\end{eqnarray}
where $\Lambda = {\rm diag}(\lambda_1,\ldots,\lambda_p)$ consists of $p$ distinct eigenvalues in descending order with the corresponding eigenvector matrix $\Gamma = [\gamma_1,\ldots,\gamma_p]$. In this article, we assume the target rank $r$ is pre-specified. An important purpose of PCA is to estimate the leading rank-$r$ eigen-subspace
\begin{eqnarray}
\Sr=\rmspan(\Gamma_r)\quad{\rm with}\quad \Gamma_r=[\gamma_1,\ldots,\gamma_r],
\end{eqnarray}
where the ordering of $\gamma_j$ is determined by the ordering of $\lambda_j$ (and hence, the equivalent wording ``ordering of eigenvalues'' and ``ordering of eigenvectors'' will be exchangeably used in the rest of discussion). Subsequent analysis can then be based on the projection of $X$ onto $\Sr$, i.e., $\Gamma_r^\top X$, without losing much information in $X$. The usual PCA starts from calculating the sample covariance matrix $\widehat S$ of the data matrix $\Xb=[X_1,\ldots,X_n]^\top$. It then performs the EVD on $\widehat S=\widetilde\Gamma \widetilde\Lambda \widetilde\Gamma^\top$, where $\widetilde\Lambda={\rm diag}(\widetilde\lambda_1,\ldots,\widetilde\lambda_p)$ consists of eigenvalues in descending order and $\widetilde\Gamma=[\widetilde\gamma_1,\ldots,\widetilde\gamma_p]$ consists of the corresponding eigenvectors, and finally it outputs $[\widetilde\gamma_1,\ldots,\widetilde\gamma_r]$ to estimate $\Sr$. Note that the ordering of $\widetilde\gamma_j$'s is determined by the ordering of $\widetilde\lambda_j$'s.
Thus, an accurate estimation of $\Sr$ depends not only on the quality of $\widetilde\gamma_j$'s, but also the correctness of the ordering of the eigenvalue estimates $\widetilde\lambda_j$'s.

\subsection{Motivating examples}

The usual PCA is known to be sensitive to the presence of outliers, and may not perform well when the normality assumption for $F$ is violated. The following example reveals one potential drawback that PCA may have in estimating $\Sr$ when the underlying data is contaminated.

\begin{ex}\label{example.pca} Assume $\Sigma=a\xi\xi^\top+(I-\xi\xi^\top)$, where $\xi$ is the leading eigenvector of PCA with signal size $a>1$, and $\Sr=\rmspan(\xi)$ with $r=1$. Consider the case, where $F$ is contaminated by a small fraction $\epsilon$ of samples along the unit vector $\nu$ satisfying $\nu^\top\xi=0$ with strength $\eta>0$. This leads to the perturbed covariance matrix for PCA to be
\begin{eqnarray*}
\Sigma^{\rm (pca)}_\epsilon = (1-\epsilon)\Sigma+\epsilon(\eta\nu\nu^\top)
   = (1-\epsilon)a\xi\xi^\top+(1-\epsilon+\epsilon\eta)\nu\nu^\top+(1-\epsilon)Q,
\end{eqnarray*}
where $Q=I-\xi\xi^\top-\nu\nu^\top$. If the noise size $\eta$ is large enough so that
\begin{eqnarray}
\eta > \frac{1-\epsilon}{\epsilon}(a-1),
\label{ex.PCA_critical}
\end{eqnarray}
then $\nu$ becomes the leading eigenvector of $\Sigma^{\rm (pca)}_\epsilon$. In this situation, $\widetilde\gamma_1$ targets $\nu$ and $\widetilde\gamma_2$ targets $\xi$. That is, PCA fails to identity $\xi$ as the leading eigenvector in estimating $\Sr$.
\end{ex}
\noindent This example shows that outliers can swap the ordering of eigenvalues even if the eigenvectors are not affected. That is, $\xi$ is still an eigenvector of $\Sigma^{\rm (pca)}_\epsilon$ but no longer the leading one. This finding conveys an important message that  maintaining the correct ordering of eigenvalues is crucial to the estimation of $\Sr$.

To improve the performance of PCA, Yata and Aoshima (2010) proposed the cross-data-matrix PCA (CDM-PCA). Different from PCA, CDM-PCA starts from randomly dividing the data $\Xb$ into two parts $\{\Xb_1,\Xb_2\}$, where $ \Xb_k=[X_{1(k)},\ldots,X_{n_k(k)}]^\top\in \mathbb{R}^{n_k\times p}, k=1,2$. For simplicity but without loss of generality, we assume $n$ is an even number and $n_1=n_2=n/2$, and assume that $\{\Xb,\Xb_1,\Xb_2\}$ are centered with zero mean. Apply singular value decomposition (SVD) to the cross-data matrix $\frac{1}{\sqrt{n_1n_2}}\Xb_1\Xb_2^\top=\bar V_1 \bar \Lambda \bar V_2^\top$, where $\{\bar V_1, \bar V_2\}$ are left and right singular vectors and $\bar \Lambda$
consists of positive singular values. Then, CDM-PCA estimates $(\Lambda,\Gamma)$ by $(\bar\Lambda, \bar\Gamma)$, where
\begin{eqnarray}\label{Gamma_bar}
\bar\Gamma=\frac{1}{2}(\bar\Gamma_1+\bar\Gamma_2)\quad{\rm with}\quad\bar\Gamma_k = \frac{1}{\sqrt{n_k}}\Xb_k^\top \bar V_k\bar\Lambda^{-\frac{1}{2}},\quad k=1,2.
\end{eqnarray}
Note that $\bar\Gamma_1^\top \bar\Gamma_2 =I$ and there is no need to align the directions of $\{\bar\Gamma_1, \bar\Gamma_2\}$ before averaging them. Under the generalized spiked model with $\lambda_j=a_j p^{\alpha_j}$ for $j\le r$ and $\lambda_j=a_j$ for $j>r$, where $a_j$ and $\alpha_j$ are some constants, and under the high-dimension-low-sample-size framework, Yata and Aoshima (2010) showed that CDM-PCA is superior to the usual PCA in identifying the leading $r$ eigenvalues. This superiority is manifested in the sense that the consistency of CDM-PCA is valid for a wider range of $\alpha_j$'s compared to PCA. Later, a comparison measure for CDM-PCA and PCA was derived by Wang and Huang (2022), which indicates that CDM-PCA is better suitable than the usual PCA for data with high-dimensionality, high noise correlation, and high noise-to-signal ratio.

A fundamental difference between the usual PCA and CDM-PCA lies in the integration of sample covariance matrices. Let $\widehat S_k$ be the sample covariance matrix of $\Xb_k$, $k=1,2$. Wang and Huang (2022) showed that $\bar\Gamma_1$ and $\bar\Gamma_2$ consist of the eigenvectors of $\widehat S_1\widehat S_2$ and $\widehat S_2\widehat S_1$, respectively, with the common eigenvalue $\bar\Lambda^2$. That is,
\begin{eqnarray}
\widehat S_1\widehat S_2 = \bar\Gamma_1 \bar\Lambda^2 \bar\Gamma_2^\top~~ {\rm with}~~ \bar\Gamma_1^\top \bar\Gamma_2=I.\label{cdmpca}
\end{eqnarray}
In other words, CDM-PCA integrates two covariance matrices $\{\widehat S_1, \widehat S_2\}$ via the product $\widehat S_1\widehat S_2$, while PCA integrates them via the average $\widehat S=\frac{1}{2}(\widehat S_1 + \widehat S_2)$. Though such integration via the average is widely used in many statistical methods, it is sensitive to the presence of outliers. The usual PCA is thus known to be a non-robust statistical method as demonstrated in Example~\ref{example.pca}. Interestingly, the use of product integration provides CDM-PCA a way to lessen the effect of outliers. The intuition behind such a robustness phenomenon is that, a signal eigenvalue will be present in both $\widehat S_1$ and $\widehat S_2$, while each single outlier instance can only appear in one of $\{\widehat S_1,\widehat S_2\}$. As a result, the size of a signal eigenvalue of $\widehat S_1\widehat S_2$ is squared, while this is not the case for an occasionally occurring outlier. Thus, the ordering of CDM-PCA's signal eigenvalues tends to be less affected by outliers than the ordering of signal eigenvalues of PCA. We use the following example to demonstrate this phenomenon.

\begin{ex} [continued from Example~\ref{example.pca}] \label{example.cpca}
Consider the same covariance $\Sigma=a\xi\xi^\top+(I-\xi\xi^\top)$ and contamination mechanism $\eta\nu\nu^\top$ in Example~\ref{example.pca}. Recall that CDM-PCA randomly splits $\Xb$ into $\{\Xb_1,\Xb_2\}$ with equal sizes. Suppose that the contaminated samples are allocated to $\Xb_1$ only. Then, the targeted covariance matrices of $\{\Xb_1,\Xb_2\}$ are
\begin{eqnarray*}
\Xb_1\to (1-2\varepsilon) \Sigma +2\varepsilon \eta\nu\nu^\top\quad{\rm and}\quad {\rm \Xb_2}\to \Sigma.
\end{eqnarray*}
Here the $2\varepsilon$ reflects the fact that $\Xb_1$ contains only half the sample size of $\Xb$ and, hence, the influence of the contamination to $\Xb_1$ is multiplied by 2 (see also (\ref{perturbation}) and the text surrounding it for more explanation). This leads to the perturbed product covariance matrix for conducting CDM-PCA to be
\begin{eqnarray*}
\Sigma^{\rm (cdmpca)}_\varepsilon =\left\{(1-2\epsilon)\Sigma+ 2\epsilon\eta\nu\nu^\top\right\}\Sigma= (1-2\epsilon)a^2\xi\xi^\top+(1-2\epsilon+2\epsilon\eta)\nu\nu^\top+(1-2\epsilon)Q.
\end{eqnarray*}
Direct calculation shows that $\nu$ becomes the leading eigenvector of $\Sigma^{\rm (cdmpca)}_\varepsilon$ if
\begin{eqnarray}
\eta > \frac{1-2\epsilon}{2\epsilon}(a^2-1).
\label{ex.CPCA_critical}
\end{eqnarray}
Comparing (\ref{ex.CPCA_critical}) with (\ref{ex.PCA_critical}), it is evident that CDM-PCA is more robust in preserving the ordering of signal eigenvectors (i.e., $\nu$ is more difficult to become the leading eigenvector) than the usual PCA if the following inequality is met,
\begin{eqnarray}\label{robust_condition}
\frac{1-2\epsilon}{2\epsilon}(a^2-1) > \frac{1-\epsilon}{\epsilon}(a-1) \quad \Longleftrightarrow \quad a>\frac1{1-2\epsilon}.
\end{eqnarray}
Recall that $a>1$, and condition~(\ref{robust_condition}) holds for sufficiently small $\epsilon$. Here we assume that individual outliers having an effect in a common direction $\nu$ is rare, and thus it is reasonable to assume that $\epsilon$ is small (and hence, the $\nu$ of an outlier will almost certainly not be simultaneously observed in both $\Xb_1$ and $\Xb_2$). This implies that there is a tendency for condition (\ref{robust_condition}) to be satisfied.
\end{ex}

Example~\ref{example.cpca} suggests a better potential for the ordering-robustness of CDM-PCA against outliers. A major aim of this article is to investigate this ordering-robustness property. To the best of our knowledge, there is no prior work in the literature on the theoretical study of ordering-robustness.

\subsection{Contribution and organization}

The original formulation of CDM-PCA in Yata and Aoshima (2010) is not convenient for the theoretical study of ordering-robustness due to its construction (\ref{Gamma_bar}) and the non-orthogonality of $\{\bar\Gamma_1, \bar\Gamma_2\}$. A modified version of CDM-PCA is proposed in Section~\ref{sec.ppca}, which we call product-PCA (PPCA). We will show that PPCA and CDM-PCA produce the same estimates of $\Lambda$. However, the counterparts of~(\ref{Gamma_bar}) obtained from PPCA are orthogonal matrices. This orthogonality of the $\Gamma$-estimates in PPCA leads to not only a more convenient framework for theoretical investigation, it also provides a more stable estimation scheme. Then, we will show that, in the presence of outliers, PPCA is more robust than PCA in preserving the ordering of signal eigenvectors, and hence, in estimating $\Sr$.

A crucial issue for a robust statistical method is its efficiency loss. Usually, there exists a trade-off between robustness and efficiency, and one might expect to suffer an efficiency loss in exchange for a robustness gain. However, the ordering-robustness, as demonstrated in Example~\ref{example.cpca}, is of a nature different from most existing robust PCA methods. It involves neither a robust loss function such as the Huber loss, nor a down-weighting scheme to mitigate the effects of outliers. The ordering-robustness potential of PPCA in Example~\ref{example.cpca} comes from the random split of the dataset and the product integration of $\{\widehat S_1, \widehat S_2\}$. It is thus of interest to investigate how much cost PPCA needs to pay in order to achieve a more robust estimation for the leading eigen-subspace $\Sr$. A surprising finding is that the ordering-robustness of PPCA can be achieved without an efficiency loss in comparison with the usual PCA. This leads to the main findings of this paper as stated below:
\begin{enumerate}
\item[(A)]
In the absence of outliers, PPCA and PCA have the same asymptotic distribution in estimating the signal eigenvalues $\{\lambda_j\}_{j\le r}$ and eigenvectors $\{\gamma_j\}_{j\le r}$. This implies that PPCA has no efficiency loss in comparison with PCA in estimating $\Sr$

\item[(B)]
In the presence of outliers, PPCA has a more stable influence function with respect to the orderings of signal eigenvectors $\{\gamma_j\}_{j\le r}$ than PCA. This implies that PPCA is more robust in estimating $\Sr$ than PCA.
\end{enumerate}
These findings suggest that PPCA has the potential to supplement the role of the usual PCA whether outliers are present or not.

The rest of this article is organized as follows. Section~\ref{sec.ppca} introduces the proposed PPCA procedure, wherein the asymptotic properties of PPCA are also derived. Section~\ref{sec.robustness} compares the robustness of PPCA and PCA in estimating $\Sr$. Numerical studies are presented in Sections~\ref{sec.simulation}-\ref{sec.data}. The paper concludes with a discussion in Section~\ref{sec.discussion}. All proofs, along with some technical lemmas, are provided in the Appendix.

\section{The Proposed Method: Product-PCA} \label{sec.ppca}

Similar to CDM-PCA, our PPCA also starts from randomly dividing the data $\Xb$ into two disjoint subsets $\{\Xb_1,\Xb_2\}$ to obtain $\{\widehat S_1,\widehat S_2\}$. Recall from Example~\ref{example.cpca} that the product of two covariance matrices has the potential to mitigate the effect of outliers. This motivates us to consider the integration of $\{\widehat S_1,\widehat S_2\}$ via the {\it product-covariance estimator}
\begin{eqnarray}
\widehat S_{12}=\widehat S_1^{\frac{1}{2}}\widehat S_2^{\frac{1}{2}},\label{S12}
\end{eqnarray}
where $\widehat S_k^{\frac{1}{2}}$ is the positive square root matrix of $\widehat S_k$.
Note that the random splitting mechanism of $\{\Xb_1,\Xb_2\}$ ensures that both
$\{\widehat S_1,\widehat S_2\}$ are estimators of $\Sigma$ and, hence, $\widehat S_{12}$ is an estimator of $\Sigma$. Recall that the SVD of $\Sigma$ is $\Gamma\Lambda\Gamma^\top$.
This suggests that the SVD of the product-covariance estimator
$\widehat S_{12}=\widehat U\widehat\Lambda \widehat V^\top$,
where $\widehat U=[\widehat u_1,\ldots,\widehat u_p]$ and $\widehat V=[\widehat v_1,\ldots,\widehat v_p]$ consist of left and right singular vectors and $\widehat \Lambda={\rm diag}(\widehat \lambda_1,\ldots,\widehat \lambda_p)$ consists of the singular values, contains information regarding $(\Lambda,\Gamma)$. We thus propose to estimate $\Lambda$ by the singular values $\widehat \Lambda$. For eigenvectors, since both the left and right singular vectors $\{\widehat U,\widehat V\}$ are estimators of $\Gamma$, we propose to integrate $\{\widehat U,\widehat V\}$ to estimate $\Gamma$ by $\widehat\Gamma=[\widehat\gamma_1,\ldots,\widehat\gamma_p]$, where
$\widehat\gamma_j$ is the first eigenvector of $(\widehat u_j\widehat u_j^\top+\widehat v_j\widehat v_j^\top)$, or equivalently,
\begin{eqnarray}
\widehat\gamma_j = \frac{\widehat u_j +\widehat v_j}{\|\widehat u_j +\widehat v_j\|},\quad j=1,\dots,p. \label{cpca.eigenvector.est}
\end{eqnarray}
Note that $\widehat u_j$ and $\widehat v_j$ are already aligned, i.e. $\widehat u_j^\top \widehat v_j\ge 0$, via $\widehat S_{12}=\widehat U\widehat\Lambda \widehat V^\top$.
Finally, a basis of $\Sr$ is estimated by $\widehat\Gamma_r=[\widehat\gamma_1,\ldots,\widehat\gamma_r]$, {where the ordering of $\widehat\gamma_j$ is determined by the ordering of $\widehat\lambda_j$. The detailed implementation procedure is summarized below. \\[-1ex]

\hrule  \vspace{0.25cm} \noindent {\bf Algorithm (Implementation of PPCA)} \vspace{0.25cm} \hrule
\begin{enumerate}
\item
Randomly split $\Xb$ into $\{\Xb_1,\Xb_2\}$ with equal size, and obtain the corresponding sample covariance matrices $\{\widehat S_1,\widehat S_2\}$.

\item
Obtain the product-covariance estimator $\widehat S_{12}=\widehat S_1^{\frac{1}{2}}\widehat S_2^{\frac{1}{2}}$ and its SVD, $\widehat S_{12} =\widehat U\widehat\Lambda \widehat V^\top$.

\item
Report the PPCA eigenvalue/eigenvector estimators $ (\widehat\Lambda, \widehat\Gamma)$, where $\widehat\gamma_j$  is the first eigenvector of $(\widehat u_j\widehat u_j^\top+\widehat v_j\widehat v_j^\top)$, equivalently, $\widehat\gamma_j=\frac{\widehat u_j +\widehat v_j}{\|\widehat u_j +\widehat v_j\|}$, $j=1,\ldots, p$.
\end{enumerate}
\hrule  \vspace{3ex}

We now state the asymptotic properties of PPCA. For comparison, the asymptotic properties of the usual PCA are also included.

\begin{thm}\label{thm.ppca_asymptotics}
Let $\beta=({\rm diag}(\Lambda)^\top,\vec(\Gamma)^\top)^\top$ and $\Sigma_\beta=H^\top WH$, where $W={\rm cov}\{\vec(XX^\top)\}$, $H=[\gamma_1\otimes\gamma_1,\ldots,\gamma_p\otimes\gamma_p, \gamma_1\otimes M_1,\ldots,\gamma_p\otimes M_p]$, and $M_j=(\lambda_jI-\Sigma)^+$, $j=1\ldots,p$. Under fixed $p$ and as $n\to\infty$, we have the following results of weak convergence:
\begin{enumerate}
\item For PPCA with $\widehat\beta=({\rm diag}(\widehat\Lambda)^\top,\vec(\widehat\Gamma)^\top)^\top$, we
have $\sqrt{n}(\widehat\beta-\beta) \stackrel{d}{\to}  N(0_{p(p+1)}, \Sigma_\beta)$.

\item For PCA with $\widetilde\beta=({\rm diag}(\widetilde\Lambda)^\top,\vec(\widetilde\Gamma)^\top)^\top$,
we have $\sqrt{n}(\widetilde\beta-\beta) \stackrel{d}{\to}  N(0_{p(p+1)}, \Sigma_\beta)$.
\end{enumerate}
\end{thm}

The extension of Theorem~\ref{thm.ppca_asymptotics} to the case of a diverging $p$ is attainable.
Consider the generalized spiked model (GSM) assumption:
\begin{eqnarray}
{\rm (GSM)}\quad \lambda_j=a_jp^{\alpha}~~{\rm for}~~ j\le r \quad{\rm and}\quad \lambda_k<\infty ~~{\rm for}~~k>r ,\label{gsm}
\end{eqnarray}
where $a_j$ and $\alpha$ are some constants that preserve the ordering of eigenvalues. For the case of PCA, Yata and Aoshima (2009), Wang and Fan (2017), and Wang {\it et al}. (2020) have shown that, for any $j\le r$, the following asymptotic properties hold:
\begin{eqnarray}
&&\sqrt{n}(\widetilde\lambda_j/\lambda_j-1)\stackrel{d}{\to}N\big(0,{\rm var}(z_j^2) \big),  \label{asymptotic_pca1}\\
&&\sqrt{n}\gamma_j^\top(\widetilde\gamma_j-\gamma_j)=o_p(1),  \label{asymptotic_pca2}\\
&& \sqrt{n}\gamma_k^\top(\widetilde\gamma_j-\gamma_j)\stackrel{d}{\to}
N\Big(0,\frac{\lambda_j\lambda_k}{(\lambda_j-\lambda_k)^2}E(z_j^2z_k^2) \Big),\quad k\le r, \quad k\ne j,\label{asymptotic_pca3}
\end{eqnarray}
where $(z_1,\dots,z_p)^\top$ is distributed as $\Sigma^{-\frac{1}{2}}X$, provided that $n\to\infty$ and $p\to\infty$ such that $p^{2(1-\alpha)}/n\to 0$. The extension for the case of PPCA is mainly based on the following result.

\begin{thm}\label{thm.ppca_asymptotics.diverge_p}
Assume GSM~(\ref{gsm}) and assume that $n\to\infty$ and $p\to\infty$ such that $p^{2(1-\alpha)}/n\to 0$. Then, we can establish that,  for any $j,k\le r$,
\begin{equation}\label{ppca_pca}
\sqrt{n}(\widehat\lambda_j-\widetilde\lambda_j)/\lambda_j=o_p(1) \quad {\rm and}\quad \sqrt{n}\gamma_k^\top(\widehat\gamma_j-\widetilde\gamma_j)=o_p(1).
\end{equation}
\end{thm}

\noindent
The above theorem states that $\{\hat\lambda_j,\hat\gamma_j\}$ of PPCA and  $\{\tilde\lambda_j,\tilde\gamma_j\}$ of PCA share the same asymptotic behavior up to the order of $o_p(n^{-1/2})$ in estimating the signal eigenvalues $\{\lambda_j\}_{j\le r}$ and eigenvectors $\{\gamma_j\}_{j\le r}$. Hence, it ensures that PPCA also exhibits the same weak convergence as PCA, as stated  in~(\ref{asymptotic_pca1})-(\ref{asymptotic_pca3}), under diverging $p$. This holds when the required conditions for $(n,p,\alpha)$ are satisfied, despite their disparate approaches to integrating $\{\widehat S_1, \widehat S_2\}$.
Consequently, from either Theorem~\ref{thm.ppca_asymptotics} for the case of finite $p$ or from Theorem~\ref{thm.ppca_asymptotics.diverge_p} for the case of diverging $p$, we conclude that using PPCA to estimate $\Sr$ incurs no efficiency loss compared to PCA. On the other hand, in Example~\ref{example.cpca}, we observe that the ordering of the leading eigenvectors of the product-covariance $\widehat S_{12}$ appears to be less affected by outliers than that of the usual sample covariance matrix $\widehat S$. This implies that PPCA has the potential to be more robust than PCA in estimating $\Sr$, without compromising estimation efficiency. We will delve into a rigorous examination of this matter in the next section.


\section{The Robustness of PPCA}\label{sec.robustness}

The aim of this section is to compare the robustness of PPCA and PCA in estimating $\mathcal{S}_r$. Note that an accurate estimation of $\mathcal{S}_r$ requires that the sample signal eigenvalues be ordered as leading eigenvalues instead of being ordered behind some noise eigenvalues. To quantitatively investigate such ordering-robustness, define
\begin{eqnarray}
\rho_{jk}=\frac{\lambda_j}{\lambda_j+\lambda_k}  \in (0.5,1),\quad  j\le r<k,  \label{rho_jk}
\end{eqnarray}
which reflects the relative size of the signal eigenvalue $\lambda_j$ with respect to its addition with noise eigenvalue $\lambda_k$. A larger value, closer to one, of $\rho_{jk}$ indicates that the sample version of $\lambda_j$ is more easily ranked correctly ahead of the sample version of $\lambda_k$. Hence, $\{\rho_{jk}\}_{k>r}$ can serve as the easiness indicator for the sample version of $\gamma_j$ being correctly included in the estimation of $\Sr$. The robustness of PPCA and PCA in estimating $\mathcal{S}_r$ can then be investigated by evaluating their robustness in estimating $\{\theta_j\}_{j\le r}$, where
\begin{eqnarray}
\theta_j = (\gamma_j^\top,\rho_{j,r+1},\ldots,\rho_{jp})^\top.\label{theta_j}
\end{eqnarray}
In particular, by treating $\theta_j$ as a functional of the underlying distribution, we will investigate its instantaneous change when the underlying distribution is perturbed by $x$. The less a statistical functional is affected by $x$, the more robustness it possesses.

\subsection{Functional representation and perturbation mechanism}\label{sec.functional_representation}

We first define the functional representation of $\theta_j$. Let $F$ be the distribution function generating $\Xb$, and let $F_k$ be the distribution function generating $\Xb_k$, $k=1,2$. Further, let $\Sigma(F)$ denote the covariance matrix as a function of the distribution $F$. The EVD in the usual PCA is expressed as
\begin{eqnarray}
\Sigma(F) &=& \Gamma(F)\cdot\Lambda(F)\cdot\{\Gamma(F)\}^\top.   \label{PCA_functional}
\end{eqnarray}
The {\it PCA-functional} $\theta_j(F)$ can be induced from $\{\Lambda(F),\Gamma(F)\}$ in~(\ref{PCA_functional}). For PPCA, the functional forms for $(\Lambda,U,V)$ have the following representation via SVD:
\begin{eqnarray}
\{\Sigma(F_1)\}^{\frac{1}{2}}\cdot\{\Sigma(F_2)\}^{\frac{1}{2}} &=& U(F_1,F_2)\cdot \Lambda(F_1,F_2) \cdot\{V(F_1,F_2)\}^\top, \label{PPCA_functional}
\end{eqnarray}
where $\Lambda(F_1,F_2)$ consists of singular values in descending order, and $U(F_1,F_2)$ and $V(F_1,F_2)$ consist of the associated left and right singular vectors. Let $\Gamma(F_1,F_2)$ be the integrated eigenvectors of $\{U(F_1,F_2), V(F_1,F_2)\}$ as given in (\ref{cpca.eigenvector.est}). The {\it PPCA-functional}\, $\theta_j(F_1,F_2)$ can be induced from $\{\Lambda(F_1,F_2),\Gamma(F_1,F_2)\}$.

To compare the robustness of the PPCA and PCA functionals, we consider a perturbation mechanism that operates under the assumption:
\begin{center}
``\it A unique outlier $x$ can affect only one of the target populations $\{F_1, F_2\}$.''
\end{center}
This assumption aligns with the practical occurrence of outliers, which may appear repeatedly but are unlikely to occur multiple times as the exact same instance. As a result, a unique outlier $x$ can belong to only one of the sub-samples $\{\Xb_1,\Xb_2\}$ generated from a random split of $\Xb$ when implementing PPCA, in this situation $x$ can influence the empirical version of either $F_1$ or $F_2$, but not both. This gives the perturbation mechanism for PPCA to be
\begin{equation}\label{perturbation.ppca}
\begin{array}{ll}
(F_1,F_2) \big|_{F_1=F_2=F} \to (F_{x,\varepsilon'},F)\quad{\rm or}\quad (F_1,F_2)\big|_{F_1=F_2=F} \to (F,F_{x,\varepsilon'}), 
\end{array}
\end{equation}
where $F_1$ and $F_2$ are both evaluated at $F$, $F_{x, \varepsilon'} = (1-\varepsilon')F + \varepsilon'\delta_x$ with a small $\varepsilon' > 0$, and $\delta_x$ is the Dirac measure at $x$. Recall that PPCA uses a random partition $\{\Xb_1,\Xb_2\}$ of $\Xb$ with equal size $n/2$, implying that $F = \frac{1}{2}(F_1 + F_2)$. The corresponding
perturbation mechanism for PCA, which operates on the average mixture of $F_1$ and $F_2$, is then given by
\begin{equation}\label{perturbation.pca}
\begin{array}{ll}
F=\frac{1}{2}(F_1+F_2)\stackrel{(\ref{perturbation.ppca})}{\to}\frac{1}{2}(F_{x,\varepsilon'}+F) = F_{x,\varepsilon'/2}.
\end{array}
\end{equation}
Setting $\varepsilon'=2\varepsilon$, we summarize the perturbation mechanism (\ref{perturbation.ppca})--(\ref{perturbation.pca}) as follows:
\begin{equation}\label{perturbation}
\begin{array}{lll}
\mbox{PPCA}&:&\quad F\to F_{x,2\varepsilon}~{\rm for~one~of}~\{F_1,F_2\},\\
\mbox{PCA}&:& \quad F\to F_{x,\varepsilon}.
\end{array}
\end{equation}
Note that the factor $2\varepsilon$ in PPCA reflects that $x$ only affects one of $\{F_1, F_2\}$, but with double the contamination rate (see also Example~\ref{example.cpca} for illustration). Since $\theta_j(F_1, F_2) = \theta_j(F_2, F_1)$, we assume, without loss of generality, that the perturbation in PPCA occurs on $F_1$. It then suffices to compare the robustness of PPCA and PCA in estimating $\Sr$ via investigating the behavior of $\theta_j(F_{x, 2\varepsilon}, F)$ and $\theta_j(F_{x, \varepsilon})$.


\begin{rmk}
Let $\widehat F$ be the empirical distribution function of $\Xb$, and let
$\widehat F_k$ be the empirical distribution function of $\Xb_k$, $k=1,2$. Then, the sample version of PPCA is $\widehat\Lambda=\Lambda(\widehat F_1, \widehat F_2)$ and $\widehat\Gamma=\Gamma(\widehat F_1, \widehat F_2)$, and the sample version of PCA is $\widetilde\Lambda=\Lambda(\widehat F)$ and $\widetilde\Gamma=\Gamma(\widehat F)$.
\end{rmk}

\subsection{Ordering-robustness of PPCA}

This subsection aims to investigate the perturbation of $\theta_j$ when the underlying distribution is perturbed by an outlier $x$ as explained in~(\ref{perturbation}). It is important to note that the perturbations of $\theta_j(F_{x,2\varepsilon},F)$ and $\theta_j(F_{x,\varepsilon})$ discussed in this section are considered in the population sense. The results derived in Theorems~\ref{thm.ppca_gamma}-\ref{thm.ppca_robustness.special} below are, therefore, independent of the sample size $n$ and are valid for any fixed but arbitrary $p$. When combined with the asymptotic results in Theorems~\ref{thm.ppca_asymptotics}-\ref{thm.ppca_asymptotics.diverge_p}, these findings enable us to conclude the efficiency-loss free ordering-robustness property of PPCA, as stated at the end of this subsection.
We first derive the results for eigenvectors.

\begin{thm}\label{thm.ppca_gamma}
Assume the perturbation in (\ref{perturbation}). For $\varepsilon$ small enough, we have
\begin{eqnarray*}\label{}
\gamma_j^\top\{\gamma_j(F_{x,2\varepsilon},F)\}&=& 1-\frac{\varepsilon^2}{2}(\gamma_j^\top x)^2x^\top M_j^2x+o(\varepsilon^2),\\
\gamma_j^\top\{\gamma_j(F_{x,\varepsilon})\}&=& 1-\frac{\varepsilon^2}{2}(\gamma_j^\top x)^2x^\top M_j^2x+o(\varepsilon^2),
\end{eqnarray*}
where $M_j=(\lambda_jI-\Sigma)^+$, $j=1,\ldots,p$.
\end{thm}
\noindent This theorem states that, in the presence of outliers, the similarity between $\gamma_j$ and its perturbed version is the same for PPCA and PCA (up to the $\varepsilon^2$ terms).
We next investigate the behavior of $\rho_{jk}(F_{x,2\varepsilon},F)$ and $\rho_{jk}(F_{x,\varepsilon})$, which can serve as measures for ability to preserve the ordering of signal eigenvectors $\{\gamma_j\}_{j\le r}$.
Recall that a large value of $\rho_{jk}$ indicates a better tendency to correctly order $\lambda_j$ ahead of $\lambda_k$. 
Let $\eta_{jk}=\rho_{jk}(1-\rho_{jk})$ and define
\begin{eqnarray}\label{tau_jk}
\tau_{jk}(x)=\frac{1}{\eta_{jk}}\Big\{\rho_{jk}(F_{x,2\varepsilon},F)-\rho_{jk}(F_{x,\varepsilon})\Big\},\quad j\le r <k.
\end{eqnarray}
The quantity $\tau_{jk}(x)$ provides a measure of the improvement of PPCA over PCA, when the underlying distribution is perturbed by $x$. The scaling factor $\eta_{jk}^{-1}$ is included to make the differences $\{\rho_{jk}(F_{x,2\varepsilon},F)-\rho_{jk}(F_{x,\varepsilon}):j\le r< k\}$ comparable over different values of $\rho_{jk}$. Define the {\it total improvement} of PPCA over PCA for the ordering of $\{\gamma_j\}_{j\le r}$ by
\begin{eqnarray}\label{tau}
\tau(x)=\frac{1}{r(p-r)}\sum_{j\le r}\sum_{k>r}\tau_{jk}(x).
\end{eqnarray}
A positive value of $\tau(x)$ indicates that PPCA is more robust to the influence of $x$ on the ordering of $\{\gamma_j\}_{j\le r}$ than the usual PCA. The following theorem is the main foundation to support the ordering-robustness of PPCA.

\begin{thm}[Ordering-robustness of PPCA]\label{thm.ppca_robustness}
Assume the perturbation in (\ref{perturbation}). For $\varepsilon$ small enough, we have the following results:
\begin{enumerate}
\item[(a)]
For any pair $(j,k)$ satisfying $j\le r<k$, we have
\begin{eqnarray*}\label{}
\rho_{jk}(F_{x,2\varepsilon},F)&=&\rho_{jk}+\varepsilon\eta_{jk}\left(d_j-d_k\right)+o(\varepsilon),\\
\rho_{jk}(F_{x,\varepsilon})&=&\rho_{jk}+\varepsilon\eta_{jk}\left(d_j-d_k\right)+o(\varepsilon),\\
\tau_{jk}(x) &=& \varepsilon^2\left\{d_kx^\top(\lambda_k I+\Sigma)^{-1}x-d_jx^\top(\lambda_j I+\Sigma)^{-1}x\right\}+ o(\varepsilon^2),
\end{eqnarray*}
where $d_j=\frac{1}{\lambda_j}(\gamma_j^\top x)^2$ is the squared standardized distance of $x$ to the distribution center (here $\mu=0$ is assumed) along the direction $\gamma_j$, $j=1,\ldots,p$.

\item[(b)]
The total improvement can be written as
\begin{eqnarray*}
\tau(x)&=& \frac{\varepsilon^2}{2}(x^\top\Sigma^{-1}x)\Big\{\Delta(x)+\Delta'(x)\Big\} + o(\varepsilon^2),
\end{eqnarray*}
where $\Delta(x)=\frac{\sum_{k>r}
d_k}{p-r}-\frac{\sum_{j\le r}d_j}{r}$
and $\Delta'(x)=\frac{p\sum_{j\le r}\sum_{k>r}\frac{\lambda_j-\lambda_k}{\lambda_j+\lambda_k}d_jd_k}{r(p-r)(x^\top\Sigma^{-1}x)}\ge 0$ for any $x$.
\end{enumerate}
\end{thm}

\noindent
Theorem~\ref{thm.ppca_robustness}(a) indicates that the perturbed versions, $\rho_{jk}(F_{x,2\varepsilon},F)$ and $\rho_{jk}(F_{x,\varepsilon})$, can be either larger or smaller than their target value $\rho_{jk}$, depending on the sign of $(d_j-d_k)$. An outlier $x$, which has larger values of $\{d_k:k>r\}$, tends to result in smaller $\rho_{jk}(F_{x,2\varepsilon},F)$ and $\rho_{jk}(F_{x,\varepsilon})$ than $\rho_{jk}$. In this situation, both PPCA and PCA can suffer from the problem of wrongly excluding signal eigenvectors and wrongly including noise eigenvectors in estimating $\Sr$. However, this wrong exclusion-inclusion phenomenon is generally more severe in PCA than in PPCA, which can be observed by noting that the improvement $\tau_{jk}(x)$ tends to be positive when, for example, $d_k\approx d_j$ or $\lambda_j\gg\lambda_k$ for $j\le r<k$. That is, $\rho_{jk}(F_{x,2\varepsilon},F)$ of PPCA tends to have a smaller negative gap from its target value $\rho_{jk}$ than $\rho_{jk}(F_{x,\varepsilon})$ of PCA does.
Theorem~\ref{thm.ppca_robustness}(b) gives the magnitude of the total improvement $\tau(x)$. As $\Delta'(\cdot)\ge 0$, we establish the inequality:
\begin{eqnarray*}
\tau(x)
&\ge&\frac{\varepsilon^2}{2}(x^\top\Sigma^{-1}x)\Delta(x)+o(\varepsilon^2),
\end{eqnarray*}
where $\Delta(x)$ plays the key role in influencing the sign of $\tau(x)$. Notably, PPCA is guaranteed to outperform PCA in preserving the ordering of $\{\gamma_j\}_{j\le r}$ for any outlier $x$ with $\Delta(x)\ge 0$.
One can confidently anticipate positive values of $\Delta(x)$ under GSM (\ref{gsm}) when $\|x\|^2=o(p^\alpha)$. The following results further provide evidence regarding the non-negative tendency of $\Delta(x)$.

\begin{cor}\label{cor.Delta}
Let $X'$ be a randomly occurring outlier having the same mean as $X$.
\begin{enumerate}
\item[(a)]
If ${\rm cov}(X')=aI$ for some $a>0$, then $E\{\Delta(X')\}>0$.

\item[(b)]
If ${\rm cov}(X')=a\Sigma$ for some $a>0$, then $E\{\Delta(X')\}=0$. If $X'$ is further assumed to be elliptically distributed and $p>2r$, then $P\{\Delta(X')> 0\}>1/2$.
\end{enumerate}
\end{cor}

\noindent This corollary indicates the non-negativity of $E\{\Delta(X')\}$ when ${\rm cov}(X')$ is a constant multiple of either $I$ or $\Sigma$. Note also that
\begin{equation*}
P\{\tau(X')>0\}\approx P\{\Delta(X')+\Delta'(X')>0\} > P\{\Delta(X')>0\}
\end{equation*}
by the positivity of $\Delta'(\cdot)$, Corollary~\ref{cor.Delta}(b) further implies that PPCA has a probability greater than $1/2$ of outperforming PCA in preserving the ordering of $\{\gamma_j\}_{j\le r}$, even under a perturbation $X'$ having a heavy-tailed elliptical distribution.

One can gain a clearer understanding of the ordering-robustness of PPCA through a perturbation that is perpendicular to the target subspace $\Sr$. In this scenario, outliers can only influence the estimation of noise eigenvalues $\{\lambda_k\}_{k>r}$ and their associated eigenvectors, more readily influencing the estimated ordering of signal eigenvectors $\{\gamma_j\}_{j\le r}$. The following results elucidate this situation.

\begin{cor}\label{thm.ppca_robustness.special}
Assume the perturbation in (\ref{perturbation}) with $x\in\Sr^{\perp}$. For $\varepsilon$ small enough, we have the following results:
\begin{enumerate}
\item[(a)]
$\rho_{jk}(F_{x,\varepsilon})<\rho_{jk}(F_{x,2\varepsilon},F)<\rho_{jk}$ for any pair $(j,k)$ satisfying $j\le r<k$.

\item[(b)]
The total improvement satisfies
\begin{eqnarray*}
\tau(x)=
\frac{\varepsilon^2}{2(p-r)}(x^\top\Sigma^{-1}x)^2+o(\varepsilon^2)>0,
\end{eqnarray*}
which indicates that PPCA outperforms PCA in preserving the ordering of $\{\gamma_j\}_{j\le r}$.
\end{enumerate}
\end{cor}

\noindent
Corollary~\ref{thm.ppca_robustness.special}(a) states that, in the presence of an outlier $x\in\Sr^{\perp}$, the perturbed signal eigenvalue $\lambda_j$ of both PCA and PPCA is susceptible to incorrect ordering with PCA being more severely affected than PPCA. Consequently, the direction $\gamma_j$ can be wrongly excluded from the leading signal eigenvectors, leading to a biased estimation of $\Sr$. Despite the adverse effect of the outlier, $\rho_{jk}(F_{x,2\varepsilon},F)>\rho_{jk}(F_{x,\varepsilon})$ indicates that PPCA is capable of producing a more accurate
ordering for $\{\gamma_j\}_{j\le r}$ than PCA. Corollary~\ref{thm.ppca_robustness.special}(b) implies that the total improvement is always positive and is proportional to $(x^\top\Sigma^{-1}x)^2$, which corresponds to the 4th power of the Mahalanobis distance of $x$ to the distribution center (here assuming $\mu=0$). The larger the size of an outlier $x$, the more advantage that
PPCA has over PCA. This facts supports the usage of PPCA especially in the
presence of an outlier $x$ with large $x^\top\Sigma^{-1}x$, in this situation PPCA wins over PCA by a considerable margin (see also Remark~\ref{rmk.tau_order} for more discussion).


%


We have shown that PPCA tends to produce a more accurate ordering for signal eigenvectors $\{\gamma_j\}_{j\le r}$ than the usual PCA. Recall also from Section~\ref{sec.ppca} that PPCA has the same asymptotic distribution as PCA in estimating $\Sr$. We have the following main conclusion on the ordering-robustness of PPCA:\\[-1ex]

\hrule  \vspace{0.25cm} \noindent {\bf The efficiency-loss free ordering-robustness of PPCA} \vspace{0.25cm} \hrule
\begin{enumerate}
\item[(A)]\label{free_robustnessA}
In the absence of outliers, PPCA exhibits no efficiency loss compared to PCA in estimating $\Sr$ in the following two senses: (i) For the case of finite $p$, PPCA and PCA share the same asymptotic distribution in estimating $(\Lambda, \Gamma)$; (ii) For the case of diverging $p$, PPCA and PCA share the same asymptotic distribution in estimating the signal eigenvalues $\{\lambda_j\}_{j\le r}$ and eigenvectors $\{\gamma_j\}_{j\le r}$ within the subspace $\Sr$.

\item[(B)]\label{free_robustnessB}
In the presence of outliers, PPCA exhibits the same eigenvector perturbation as PCA, but is more robust than PCA in preserving the ordering of $\{\gamma_j\}_{j\le r}$ for any outlier $x$ satisfying $\tau(x)>0$, which is a measure of improvement.
\end{enumerate}
\hrule
\vspace{0.7cm}

We conclude this section by emphasizing that PPCA does not employ a down-weighting scheme to mitigate the effect of outliers. The gain in ordering-robustness for PPCA mainly arises from the random partition and the product integration $\widehat S_{12}$ of $\{\widehat S_1,\widehat S_2\}$. PPCA's implementation involves no additional tuning parameter, which is often required in many robust statistical methods to balance efficiency and robustness. This no-efficiency-loss phenomenon represents a fundamental difference between PPCA and other robust PCA methods (e.g., PCA based on robust M-estimators for the covariance matrix). The finding that PPCA and PCA share the same asymptotic distribution supports this point, as most other robust methods suffer from efficiency loss.
As a consequence, PPCA should not be expected to exhibit comparable robustness to existing robust PCA methods. Recall from Theorem~\ref{thm.ppca_robustness} that the ordering-robustness of PPCA is established for small $\varepsilon$. The requirement of a ``small $\varepsilon$'' is crucial for ordering-robustness because, in this situation, an individual outlier cannot simultaneously affect both $\{\widehat S_1, \widehat S_2\}$ and therefore has less influence on PPCA than on PCA (see Example~\ref{example.cpca}). It is still possible for PPCA to fail when the outlier proportion is large, while robust PCA methods (with proper tuning and at the cost of efficiency loss) can deliver satisfactory performance under heavy outliers. This point, however, does not diminish the merits of PPCA, as our objective here is not to propose a new robust PCA method but to demonstrate the superiority of PPCA over the usual PCA, as summarized in (A)-(B).


\begin{rmk}\label{rmk.tau_order}
From Corollary~\ref{thm.ppca_robustness.special}(b), the magnitude of $\tau(x)$ is $O\left\{(\varepsilon^2 p)( x^\top\Sigma^{-1}x/p)^2\right\}$ and it becomes non-negligible when $x^\top\Sigma^{-1}x /p$, representing the squared Mahalanobis distance scaled by the variable dimension, has an order of magnitude $O\{1/(\varepsilon\sqrt{p})\}$ or larger. Notably, an outlier $x$ with a substantial value of $x^\top\Sigma^{-1}x/p$ tends to put forth a more pronounced influence on the ordering of signal eigenvectors. This observation highlights the significant advantage of PPCA over PCA in the presence of influential outliers.
\end{rmk}

\section{Simulation Studies}\label{sec.simulation}

\subsection{Simulation settings}

For each simulation run, the eigenvectors $\Gamma$ are randomly generated by orthogonalizing a $p\times p$ random matrix with independent $N(0,1)$ elements. The signal eigenvalues are set to be $\lambda_j=1+(p/n)^{1/2}+p^{1/(1+j)}$, $j\le r$, and the noise eigenvalues $\{\lambda_j\}_{j>r}$ are generated from $U(0.5,1.5)$. Given $(\Lambda,\Gamma)$, the data $\{X_i\}_{i=1}^n$ are generated from the mixture
\begin{eqnarray}
(1-\pi)t_\nu(0,\Sigma)+\pi t_3(\mu_{\rm out}, \Sigma_{\rm out}),\label{sim.model}
\end{eqnarray}
where $t_\nu(0,\Sigma)$ is the multivariate $t$-distribution with degrees of freedom $\nu$, mean $0$, and covariance $\Sigma$, and $\pi$ is the contamination proportion. There are two types of outliers involved in~(\ref{sim.model}):
\begin{itemize}
\item
The {\it heavy-tailed outliers} from $t_\nu(0,\Sigma)$. A smaller $\nu$ gives more heavy-tailed outliers.

\item
The {\it heterogeneous outliers} from $t_3(\mu_{\rm out}, \Sigma_{\rm out})$, where we set $\mu_{\rm out}=(n^\frac{1}{2}p^{\frac{1}{4}})^c\frac{\xi}{\|\xi\|}$ as motivated by Remark~\ref{rmk.tau_order} (when $\varepsilon=n^{-1}$) with $\xi\sim N(0,I)$ and $\Sigma_{\rm out}=\|\mu_{\rm out}\|^2I$, and $c>0$ controls the influence of outliers. A larger $\pi$ gives more heterogeneous outliers.
\end{itemize}
Note that the presence of heavy-tailed outliers will not affect the consistency of $\widehat S_{12}$ and $\widehat S$, but can make their estimation unstable. The presence of heterogeneous outliers will make $\widehat S_{12}$ and $\widehat S$ biased, and hence, have more influence on the performance of PPCA and PCA than heavy-tailed outliers will have. We implement PPCA, CDM-PCA, and PCA to compare their performance in estimating $\Sr$. The performance of each method is based on the similarity measure
$\xi_q=\frac{1}{r}\sum_{j=1}^{r}s_{qj}$, $q\ge r$,
where $s_{qj}$ is the $j^\th$ singular value of $B^\top\Gamma_r$, and $B$ is an orthonormal basis comprising
the leading $q$ eigenvectors obtained from PPCA, CDM-PCA, or PCA. A larger value of $\xi_q\in [0,1]$ indicates a better performance for $B$ in recovering $\Sr$, where $\xi_q=1$ indicates that $\Sr\subseteq\rmspan(B)$. The means of $\xi_q$ for $q\in\{r,r+1,\ldots, 40\}$, based on 200 replicates, are reported under the experimental settings $r=5$, $n=500$, $c\in\{0.25, 0.5, 1\}$, $p\in \{250, 1000\}$, $\nu\in\{5, 20\}$, and $\pi\in \{0,0.05\}$.

\subsection{Simulation results}

In the left two columns of Figure~\ref{fig.sim}, simulation results under $p=250$, $c=1$, and different combinations of $(\nu,\pi)$ are reported. For the simplest case $(\nu,\pi)=(20,0)$ (i.e., approximately Gaussian and with no heterogeneous outliers), one can see that PPCA and PCA have almost the same $\xi_q$ values. This supports our theoretical finding that, in the absence of outliers, both methods share the same asymptotic distributions and, hence, using PPCA to estimate $\Sr$ has no efficiency loss in comparison with PCA.
For the case of $(\nu,\pi)=(5,0)$ (i.e., with heavy-tailed outliers), PPCA is found to outperform PCA when $q$ is small, and both methods achieve similar $\xi_q$ values for $q\ge 20$. This result supports the ordering-robustness of PPCA, indicating that PPCA tends to produce a more accurate ordering of $\{\gamma_j\}_{j\le r}$ than the usual PCA does. PCA requires more eigenvectors to encompass the target leading eigen-subspace $\Sr$.
The ordering-robustness of PPCA is more evident in the case of $(\nu,\pi)=(20,0.05)$ (i.e., with heterogeneous outliers). Clearly, the heterogeneous outliers tend to have a larger Mahalanobis distance $(x^\top\Sigma^{-1}x)^{1/2}$ than heavy-tailed outliers. The simulation result confirms our finding in Corollary~\ref{thm.ppca_robustness.special} that the total improvement $\tau(x)$ is shown to be proportional to $(x^\top\Sigma^{-1}x)^2$, and PPCA shows a substantial improvement over PCA for $q<30$. When $q\ge 30$,  it becomes evident that PCA achieves the same performance as PPCA. It underscores the fact that, while both methods have the same efficiency in estimating $(\Lambda, \Gamma)$, PCA produces a worse ordering of signal eigenvectors than PPCA.
As a result, PPCA is able to use fewer eigenvectors than PCA to encompass $\Sr$. The superiority of PPCA over PCA becomes even more  apparent in the most severe case $(\nu,\pi)=(5,0.05)$ (i.e., with both heavy-tailed and heterogeneous outliers). In this scenario, PPCA consistently outperforms PCA for all values of $q$. Specifically, PPCA achieves $\xi_q=0.9$ for $q\ge 30$, while PCA fails to reach a $\xi_q$ value of $0.9$ even when $q=40$.

In the right two columns of Figure~\ref{fig.sim}, simulation results under $p=1000$, $c=1$, and different combinations of $(\nu,\pi)$ are reported. Note that in this situation, where the number of variables $p=1000$ significantly exceeds the sample size, the effect of outliers becomes more severe than in the case of $p=250$. The robustness of PPCA over PCA is clearly observed in the simulation results, and similar conclusions can be drawn for PPCA as those observed in the setting of $p=250$. Under the no-outlier case $(\nu,\pi)=(20,0)$, PPCA and PCA have similar performance. In all other cases of $(\nu,\pi)$, PPCA outperforms PCA especially for small $q$ values, and the magnitude of improvement of PPCA over PCA is larger than in the case of $p=250$. Moreover, PPCA appears to dominate the performance of PCA in the most severe case of $(\nu,\pi)=(5,0.05)$, where PPCA achieves $\xi_q=0.8$ when $q=30$, while PCA has a $\xi_q$ value no larger than $0.7$ even when $q=40$.

Upon comparing PPCA and CDM-PCA, it can be observed that PPCA and CDM-PCA exhibit nearly identical performance in estimating $\Sr$ in the absence of outliers, with $(\nu,\pi)=(20,0)$. This aligns with the observations of Wang et al. (2020), who noted that CDM-PCA and PCA share the same asymptotic distributions in estimating signal eigenvalues and their associated eigenvectors under the GSM~(\ref{gsm}). The disparity between PPCA and CDM-PCA becomes pronounced in the presence of outliers, where PPCA demonstrates better robustness in estimating $\Sr$, particularly in the most challenging scenario with $(\nu,\pi)=(5,0.05)$. Furthermore, the superiority of PPCA over CDM-PCA becomes more pronounced in high-dimensional settings with $p=1000$. These findings suggest that, while both PPCA and CDM-PCA exhibit the efficiency-loss-free property in estimating $\Sr$, PPCA tends to be more robust, especially in situations with a large $p$.

Figure~\ref{fig.sim2} presents simulation results under $\pi=0.05$ with various combinations of $(\nu,c)$. This allows us to observe the impact of $c$ on the performance of all methods, where a larger $c$ implies further outlyingness. These results are compared with the simulation results in Figure~\ref{fig.sim}, where $\pi=0.05$ and $c=1$.
As anticipated, the advantage of PPCA over PCA diminishes as $c$ decreases, consistent with our theoretical finding that  $\tau(x)\propto (x^\top\Sigma^{-1}x)^2$. Specifically, when $c =0.25$ and $0.5$, the overall improvement $\tau(x)$ is negligible, as discussed in Remark~\ref{rmk.tau_order}. Nevertheless, it is evident that PPCA consistently outperforms PCA in estimating $\Sr$ even when $c=0.25$, particularly in the scenario of $p=1000$. Furthermore, despite the diminished improvement due to the small $c$ value, PPCA still demonstrates superior performance compared to CDM-PCA. These results affirm the robustness of PPCA over PCA and CDM-PCA, even when the outliers' strength is relatively weak.

In conclusion, our simulation results reveal that PPCA is asymptotically equivalent to PCA in estimating $\Sr$ when there are no outliers. Notably, PPCA demonstrates substantial superiority over PCA and CDM-PCA in the presence of outliers. Especially for high-dimensional datasets, where $p>n$, we strongly recommend the use of PPCA.

\begin{figure}[!ht]
\hspace{-1.7cm}
\begin{center}
\includegraphics[width=2.7in]{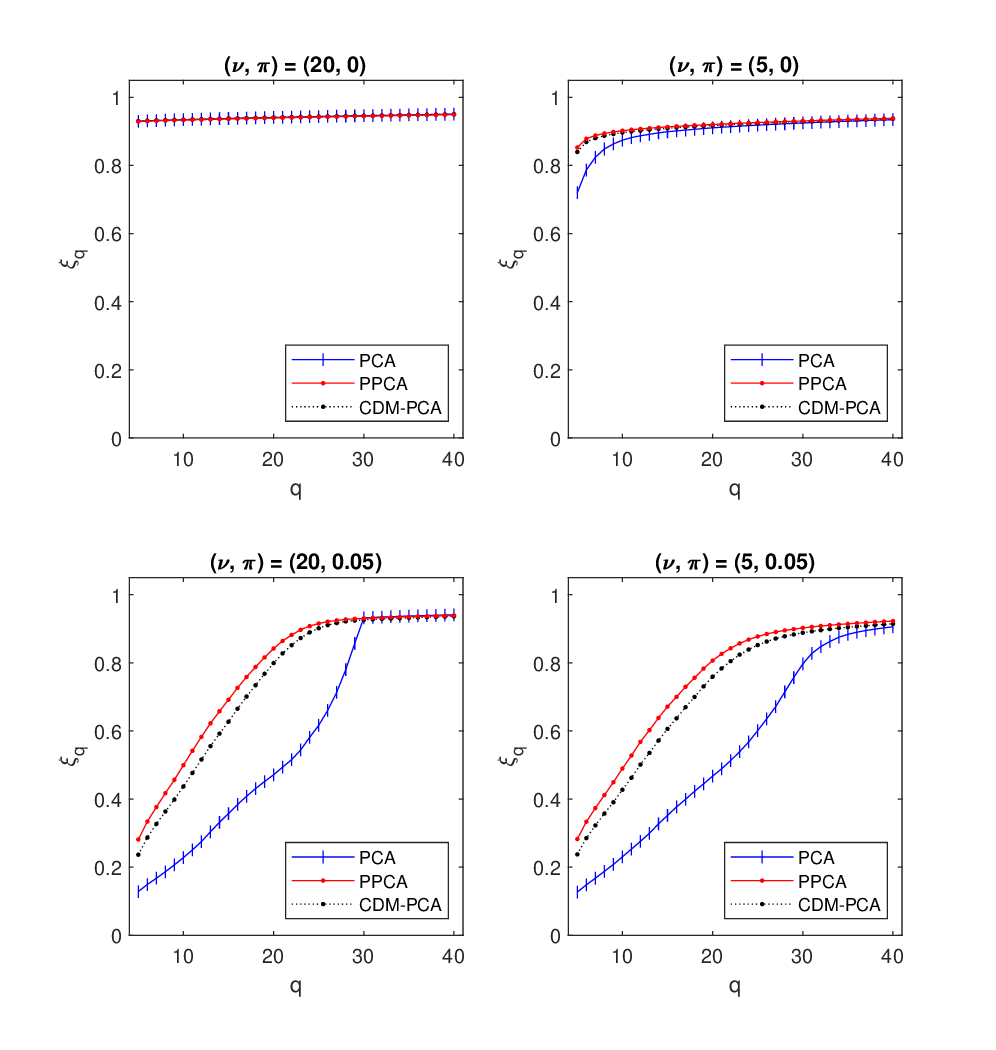}
\includegraphics[width=2.7in]{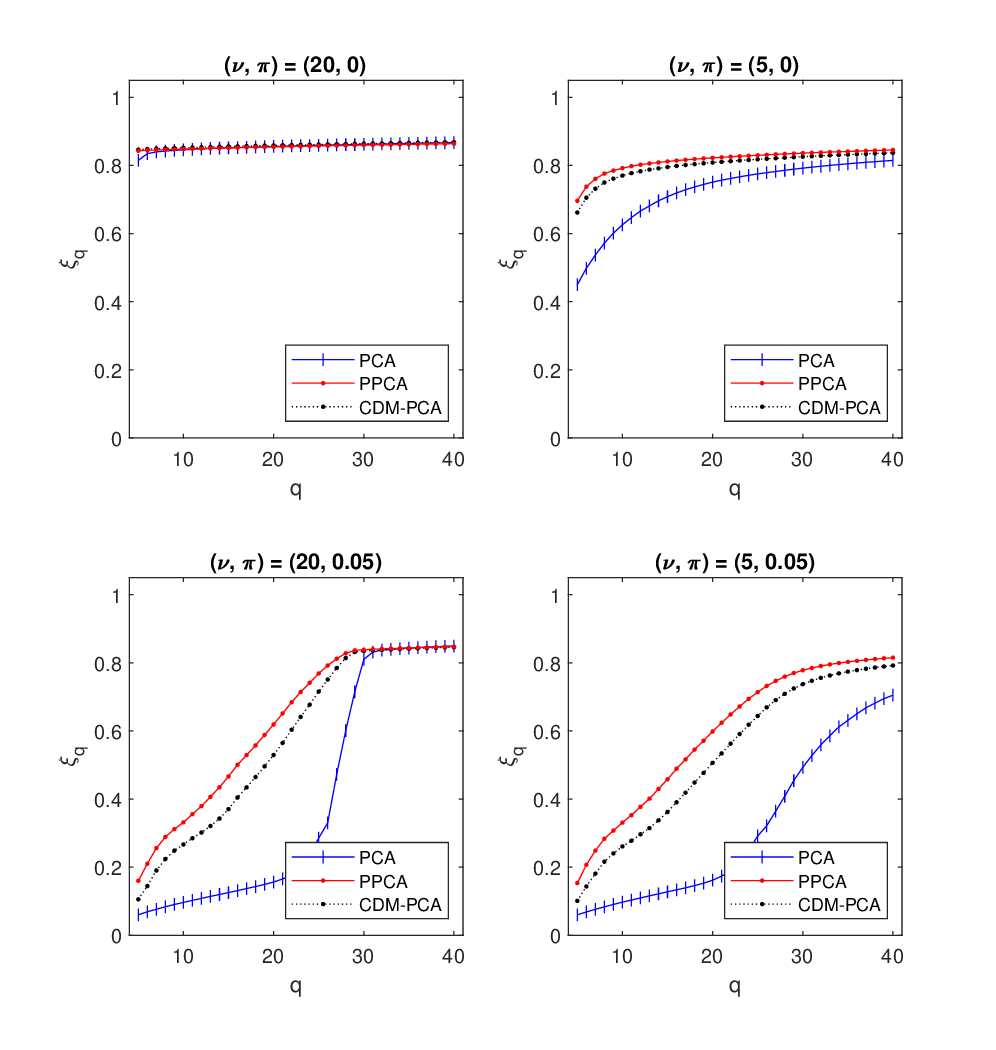}
\end{center}
\vspace{-0.5cm}
\caption{The means of the similarity measure $\xi_q$, $q\in\{r,r+1,\ldots, 40\}$, for PCA, PPCA, and CDM-PCA under $r=5$, $n=500$, $c=1$, and different combinations of $\nu\in\{5, 20\}$ and $\pi\in \{0,0.05\}$. The left two columns are for the case of $p=250$, and the right two columns are for the case of $p=1000$.}\label{fig.sim}
\end{figure}

\begin{figure}[!ht]
\hspace{-1.7cm}
\begin{center}
\includegraphics[width=2.7in]{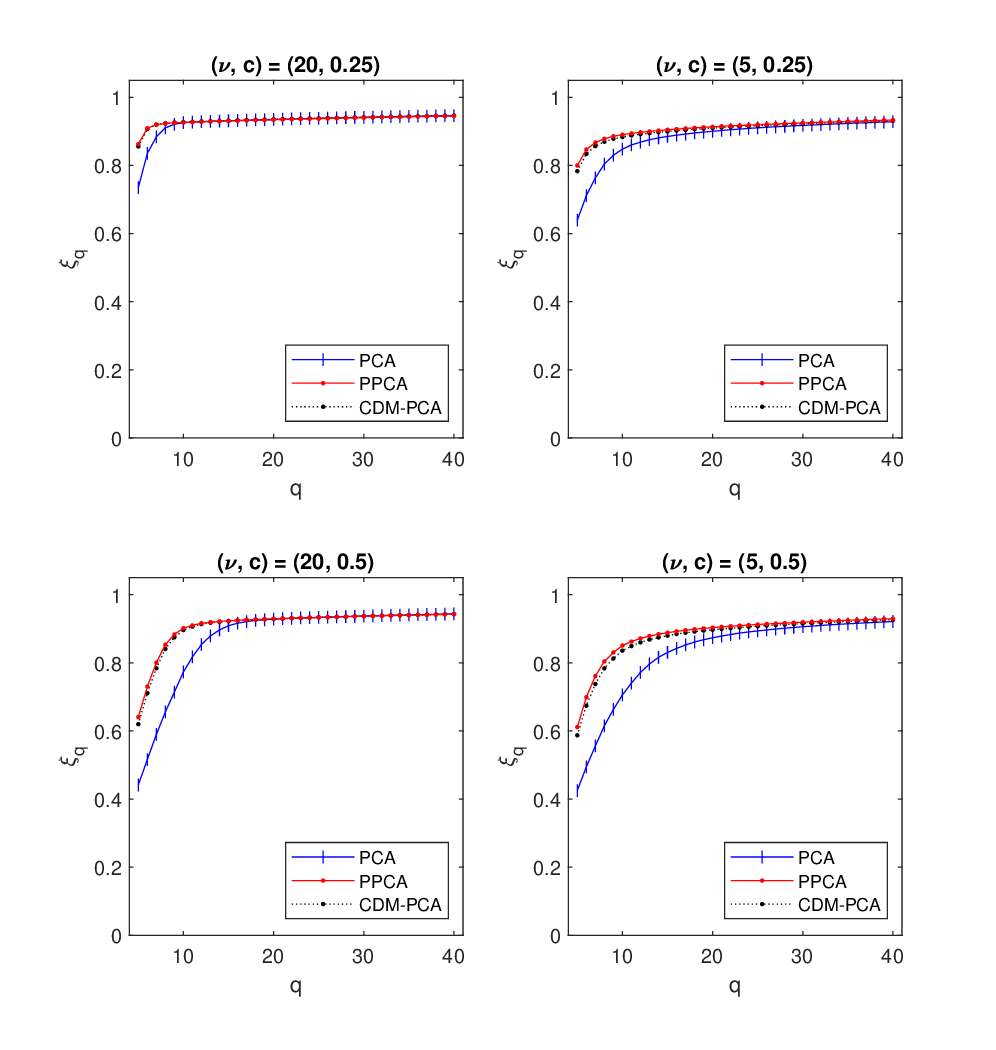}
\includegraphics[width=2.7in]{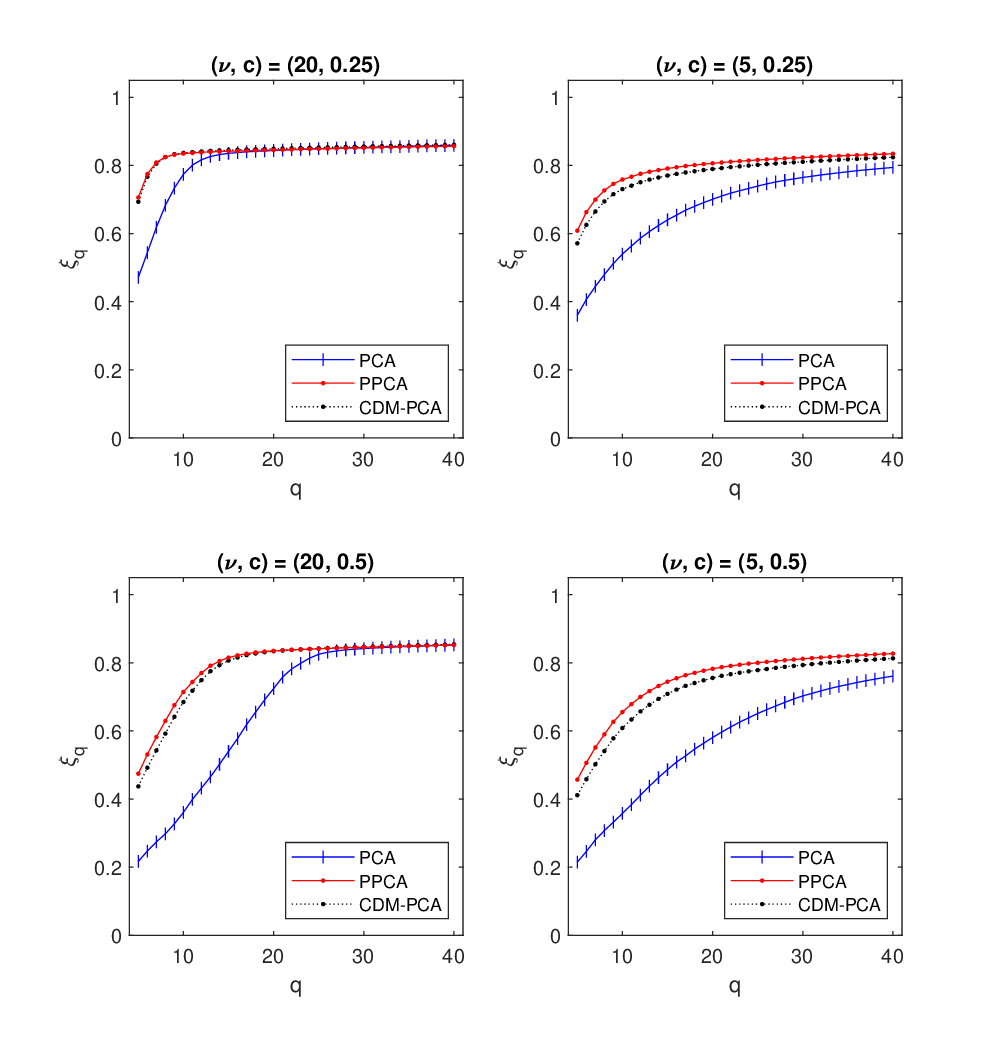}
\end{center}
\vspace{-0.5cm}
\caption{The means of the similarity measure $\xi_q$, $q\in\{r,r+1,\ldots, 40\}$, for PCA, PPCA, and CDM-PCA under $r=5$, $n=500$, $\pi=0.05$, and different combinations of $\nu\in\{5, 20\}$ and $c\in \{0.25,0.5\}$. The left two columns are for the case of $p=250$, and the right two columns are for the case of $p=1000$.}\label{fig.sim2}
\end{figure}

\section{Olivetti Faces Dataset}\label{sec.data}

The Olivetti dataset contains a set of face images taken between April 1992 and April 1994 at AT\&T Laboratories Cambridge. The dataset $\{Y_i\}_{i=1}^{400}$ contains $400$ face image matrices of size $64\times 64$ with grey levels in $[0, 255]$ from 40 people (10 images of each person). These images were taken at different times with varying lighting, facial expressions and facial details.
This dataset can be downloaded from {https://scikit-learn.org/stable/modules/generated/sklearn.datasets.fetch\_olivetti\_faces.html}.
In our analysis, we artificially contaminate a portion of the images to obtain $\Xb=[X_1,\ldots,X_n]^\top$, where $X_i=\vec(Y_i)$ is the vectorized image of size $p=64^2$. Two types of contamination mechanisms are considered: (i) 20 images are randomly drawn from the 400 images, and are added with random noises generated from $t_5(0_p,50I)$; (ii) 20 noisy images are added to the dataset, where the pixel values of each noise image are generated from the discrete uniform distribution over $\{0,1,\dots, 255\}$. Based on the resulting contaminated data $\Xb$ with $(n,p)=(420,64^2)$, we implement PCA, PPCA, and CDM-PCA to estimate $\Sr$, and then report the lower-dimensional reconstructed faces
\begin{eqnarray}
\widehat X_i = \widehat\mu + P_{\widehat{\cal S}_r}(X_i- \widehat \mu),\label{reconstructed_image}
\end{eqnarray}
where $\widehat\mu$ is the mean face of $\{X_i\}_{i=1}^n$, and $P_{\widehat{\cal S}_r}$ is the projection onto the leading rank-$r$ eigen-subspace estimated by different methods. Figure~\ref{fig.face} presents reconstructed images for some randomly drawn uncontaminated images, where the number of components is set to $r\in\{10, 20, 30, 40, 50\}$, denoted by PCA$(r)$, PPCA$(r)$, and CDM-PCA$(r)$. The mean face and the original true faces are also reported for comparison.

From Figure~\ref{fig.face}, one can clearly observe the progression of PPCA$(r)$ as $r$ increases from 10 to 30, indicating that the leading eigenvectors identified by PPCA play important roles for face reconstruction. Taking Figure~\ref{fig.face}(a) as an example, one can observe that the man turns his eyes to the right in PPCA(20) and smiles in PPCA(30). As $r$ increases from 30 to 50, while PPCA$(r)$ becomes more refined to match the true face, there seems to be no significant difference between PPCA$(30)$ and PPCA$(50)$. This observation indicates that PPCA can effectively identify important face bases as leading eigenvectors in the presence of contaminated images. In contrast, the situation is different for PCA, where one can hardly observe any obvious progress for PCA$(r)$ until $r\ge 40$. This suggests that a large portion of the leading eigenvectors of PCA tends to be noise eigenvectors and is not beneficial for face reconstruction. Moreover, it is found that PPCA$(50)$ significantly outperforms PCA$(50)$, further supporting that PCA is less robust against the presence of contaminated images as PPCA.

While a broad comparison suggests similar performance between PPCA and CDM-PCA, a closer examination reveals that PPCA consistently generates more accurate reconstructed images compared to CDM-PCA. As illustrated in Figure~\ref{fig.face}(b), PPCA(20) exhibits a clearer pattern of white teeth than CDM-PCA(20). Furthermore, PPCA(50) consistently produces less-contaminated images than CDM-PCA(50) in both cases (a) and (b). These observations align with our simulation results, showing that PPCA has better robustness than CDM-PCA.

\begin{figure}[!ht]
\begin{center}
(a)\\  \includegraphics[width=6in]{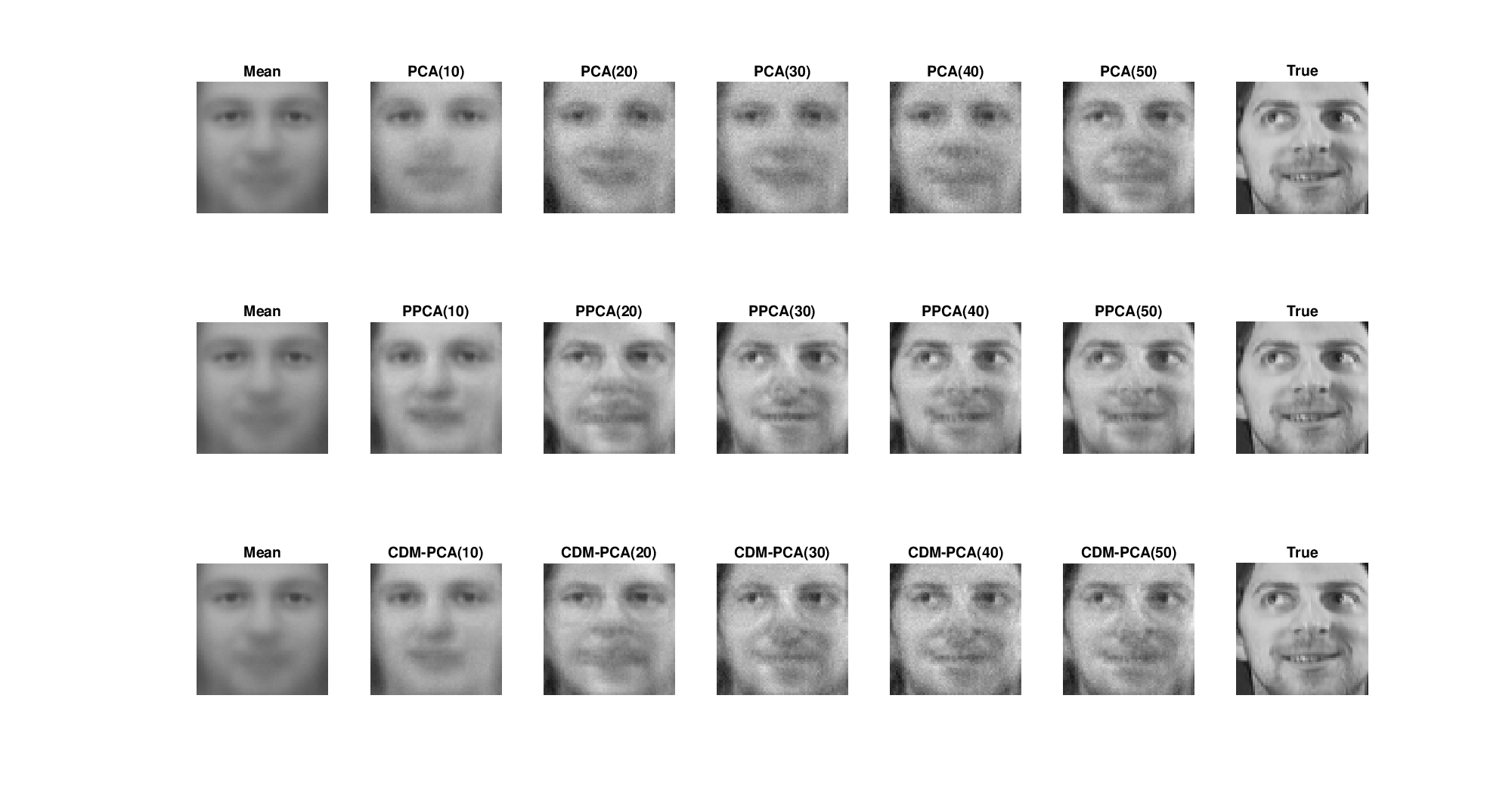}\\
(b)\\ \includegraphics[width=6in]{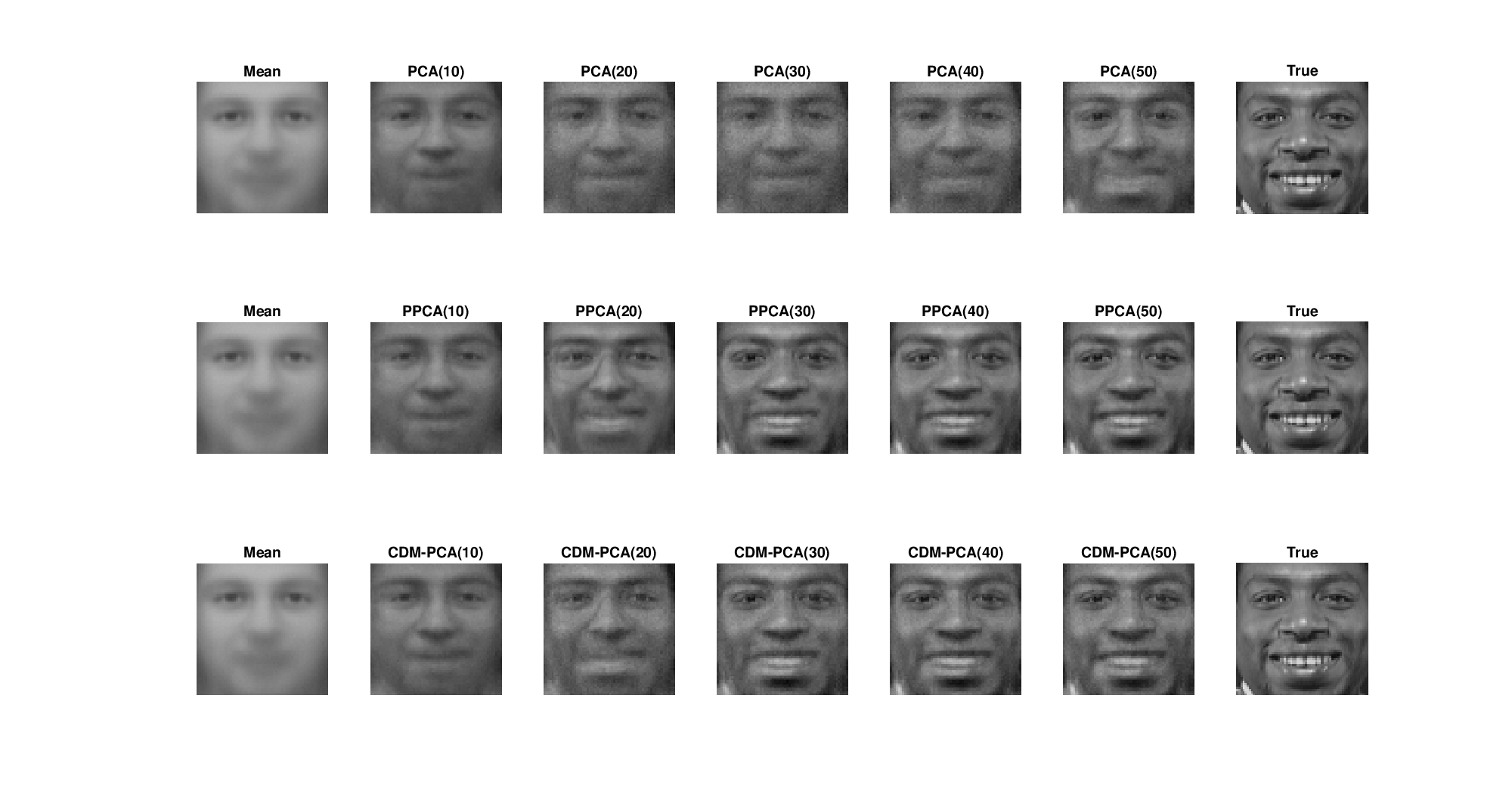}
\end{center}
\caption{Two cases of face reconstruction by PCA (rows 1\&4), PPCA (rows 2\&5), and CDM-PCA (rows 3\&6). The number inside the parentheses is the number of principal components used for reconstruction. The mean face (column 1) and the original true face (column 7) are also presented for comparison. }\label{fig.face}
\end{figure}


\section{Discussion}\label{sec.discussion}

In this article, we have modified CDM-PCA to propose a novel PPCA for the purpose of unsupervised dimension reduction. We have shown that PPCA has the potential to replace the usual PCA in the sense that (1) PPCA has no efficiency loss in estimating $\Sr$ in comparison with PCA, and (2) PPCA is more robust in preserving the right ordering of the signal eigenvectors $\{\gamma_j\}_{j\le r}$ than PCA does.
A few remarks concerning further development of PPCA are listed below.
\begin{itemize}
\item {\bf Rank estimation.}
The target rank $r$ is assumed to be known in this work. However, in practice, $r$ needs to be determined by the data. There are many methods developed for PCA rank estimation. The extension of existing rank estimation methods to PPCA worths further development.


\item
{\bf Comparison with CDM-PCA.}
Both PPCA and CDM-PCA are grounded in the random partition into two subsets and the product integration of $\{\widehat S_1, \widehat S_2\}$. PPCA employs SVD on $\widehat S_1^{\frac{1}{2}}\widehat S_2^{\frac{1}{2}}$ to extract singular vectors $\{\widehat U,\widehat V\}$, while CDM-PCA utilizes EVD on $\widehat S_1\widehat S_2$ to extract eigenvectors $\{\bar\Gamma_1,\bar\Gamma_2\}$. In theory, they target the same eigen-subspace and the same eigenvalues, exhibiting identical asymptotic normality. However, our numerical experience reveals that PPCA tends to be more robust than CDM-PCA in estimating $\Sr$. This observation prompts an exploration of the robustness of using the power $\frac{1}{2}$ vs. 1 and the use of SVD vs. EVD.
Note also that CDM-PCA integrates $\{\bar\Gamma_1,\bar\Gamma_2\}$ via $\frac{1}{2}(\bar\Gamma_1+\bar\Gamma_2)$, while PPCA integrates $\{\widehat U,\widehat V\}$ via $(\widehat u_j + \widehat v_j)/\|\widehat u_j + \widehat v_j\|$. These differences are implementation details and can be easily modified. Although CDM-PCA might be less robust than PPCA, its inspiring merits of using random partition and product integration should be well-recognized.

\item
{\bf High-dimensionality with weak spikes.}
PPCA and PCA are shown to share the same asymptotic properties in consistently estimating $\Sr$ under the generalized spiked model with $p\to\infty$ and $\lambda_r\to\infty$ (i.e., high-dimensionality with strong spikes). However, it is well-known that PCA is not consistent under the more challenging scenario of  $p\to\infty$ and $\lambda_1<\infty$ (i.e., high-dimensionality with weak spikes). 
This issue is rigorously addressed in Ding (2021), Bao \textit{et al.} (2022), and Ding, Li, and Yang (2024), where the authors applied random matrix theory to derive the convergence limits and asymptotic distributions of PCA. These papers demonstrate how working with asymptotic distributions can help correct PCA’s inconsistency. Relevant results are also presented in Ding (2020) and Bao, Ding, and Wang (2021) for high-dimensional matrix denoising models with finite signal sizes.
These findings highlight the importance of investigating the asymptotic properties of PPCA under the case of weak spikes. Such an analysis could provide deeper insights into the distinctions between PPCA and PCA in challenging high-dimensional regimes, and it represents an important direction for future research.
\end{itemize}



\begin{thebibliography}{}
\bibitem{}
Anderson, T. W. (1963). Asymptotic theory for principal component analysis.
{\it Annals of Mathematical Statistics}, 34(1), 122-148.


\bibitem{}
Bao, Z., Ding, X., and Wang, A. K. (2021). Singular vector and singular subspace distribution for the matrix denoising model.
{\it The Annals of Statistics}, 49(1), 370-392.

\bibitem{}
Bao, Z., Ding, X., Wang, J., and Wang, K. (2022). Statistical inference for principal components of spiked covariance matrices. {\it The Annals of Statistics}, 50(2), 1144-1169.

\color{black}
\bibitem{}
Critchley, F. (1985). Influence in principal components analysis. {\it Biometrika}, 72(3), 627-636.


\bibitem{}
Croux, C., Haesbroeck, G., and Joossens, K. (2008). Logistic discrimination using robust estimators: an influence function approach. {\it Canadian Journal of Statistics}, 36(1), 157-174.


\bibitem{}
Ding, X. (2020). High dimensional deformed rectangular matrices with applications in matrix denoising.  {\it Bernoulli}, 26(1), 387-417.

\bibitem{}
Ding, X. (2021). Spiked sample covariance matrices with possibly multiple bulk components. {\it Random Matrices: Theory and Applications}, 10(01), 2150014.
  
   
\bibitem{}
Ding, X., Li, Y., and Yang, F. (2024). Eigenvector distributions and optimal shrinkage estimators for large covariance and precision matrices. {\it arXiv preprint arXiv:2404.14751}.

\color{black}
\bibitem{}
Fernholz, L. T. (2001). On multivariate higher order von Mises expansions. {\it Metrika}, 53(2), 123-140.


\bibitem{}
Jolliffe, I. T. and Cadima, J. (2016). Principal component analysis: a review and recent developments.
{\it Philosophical Transactions of the Royal Society A: Mathematical, Physical and Engineering Sciences}, 374(2065), 20150202.

\bibitem{}
Pires, A. M. and Branco, J. A. (2002). Partial influence functions. {\it Journal of Multivariate Analysis}, 83(2), 451-468.

\bibitem{}
Tyler, D. E. (1981). Asymptotic inference for eigenvectors. {\it Annals of Statistics}, 9(4), 725-736.


\bibitem{}
Wang, S. H., Huang, S. Y., and Chen, T. L. (2020). On asymptotic normality of cross data matrix-based PCA in high dimension low sample size. {\it Journal of Multivariate Analysis}, 175, 104556.


\bibitem{}
Wang, S. H. and Huang, S. Y. (2022). Perturbation theory for cross data matrix-based PCA. {\it Journal of Multivariate Analysis}, 190, 104960.

\bibitem{}
Wang, W. and Fan, J. (2017). Asymptotics of empirical eigenstructure for high dimensional spiked covariance. {\it Annals of Statistics}, 45(3), 1342.


\bibitem{}
Yata, K. and Aoshima, M. (2009).
PCA consistency for non-Gaussian data in high dimension, low sample size context.
{\it Communications in Statistics-Theory and Methods}, 38, 2634-2652.

\bibitem{}
Yata, K. and Aoshima, M. (2010). Effective PCA for high-dimension, low-sample-size data with singular value decomposition of cross data matrix. {\it Journal of Multivariate Analysis}, 101(9), 2060-2077.
\end{thebibliography}
\end{document}